\documentclass[journal=jacsat,manuscript=article]{achemso}

\usepackage[version=3]{mhchem} 
\usepackage{multicol}
\usepackage{systeme}
\usepackage{booktabs}

\usepackage{multirow}
\usepackage{soul}
\usepackage{color}
\usepackage{siunitx} 
\usepackage[hidelinks]{hyperref}

\newcommand{\FU}{Department of Physics, Freie Universität Berlin, 14195 Berlin, Germany}

\author{Anna Wroblewska}
\affiliation[Warsaw]{Faculty of Physics, Warsaw University of Technology, Koszykowa 75, 00-662, Warsaw, Poland}
\email{Anna.Wroblewska@pw.edu.pl}
\author{Niclas S. Mueller}
\affiliation[]{\FU{}}
\author{Mariusz Zdrojek}
\affiliation[Warsaw]{Faculty of Physics, Warsaw University of Technology, Koszykowa 75, 00-662, Warsaw, Poland}

\author{Stephanie Reich }

\affiliation[FU Berlin]
{\FU{}}
\author{Georgy Gordeev}
\affiliation[FU Berlin]
{\FU{}}
\alsoaffiliation[Second University]
{Department of Physics and Materials Science, University of Luxembourg, L-4422 Belvaux, Luxembourg}
\email{georgy.gordeev@uni.lu}

 \title{Cross-Dimensional Exciton Coupling in Quantum Dot–Carbon Nanotube Hybrid Thin Films}

\begin{document}

\newcommand{\wn}[1]{\SI{#1}{\per \cm}}


\begin{abstract}


Dipole–dipole interactions provide a route to couple excitons across materials of different dimensionality. Here, we introduce hybrid films as a cross-dimensional excitonic platform and develop a theory of resonant Raman scattering in the presence of dipolar exciton coupling. Combining a coupled-oscillator model with perturbation theory, we show that coupling renormalizes the exciton–photon matrix elements and modifies nanotube Raman excitation profiles. We test this framework in vacuum-filtered films containing InP/ZnS quantum dots and mixed chirality metallic single-walled carbon nanotubes. The quantum-dot exciton remains near 1.83 eV, while chirality-dependent nanotube excitons span 1.7–\SI{1.93}{eV}, enabling systematic control of excitonic detuning. Relative to pristine nanotube films, the hybrids exhibit detuning-dependent redshifts and blueshifts, Raman-intensity enhancement, reduced effective linewidths, and near-resonant two-branch profiles. The model yields an effective splitting of approximately 110 meV. Resonant Raman scattering thus provides component-selective access to dipolar interactions that are only weakly apparent in ensemble absorption spectra.\par


\end{abstract}





\par \vspace{\baselineskip}
\noindent
Resonant dipole–dipole interactions between optical excitations can generate collective emission, energy transfer, and hybridized excitonic states, providing an important mechanism for controlling light–matter interactions at the nanoscale.\cite{Solano2017, Martn-Cano2010, Burgess2025} Cavities can strongly enhance dipole coupling, although traditional flat mirror cavities have large mode volumes,\cite{Kristensen2014} demand precise alignment, and by design favor two-dimensional materials. Although plasmonic cavities achieve smaller mode volumes,\cite{Maier2006, Russell2012} they often suffer from high damping from metallic nanoparticles. A promising alternative is cavity-free dipole-dipole coupling realized by arranging closely packed dipoles into a crystal\cite{Munn1988}, combining low damping with small effective mode volumes. Such dipolar crystals may consist of various dipole types: plasmonic lattices can attain high light-matter coupling regimes\cite{Mueller2020,Boddeti2022} despite losses, molecular 3D assemblies show exciton hybridization and Davydov splitting\cite{Davydov1964,Nematiaram2021,Pandya2021}, and low-dimensional dipolar crystals yield collective excitonic states with superradiant emission.\cite{Juergensen2023} Extending this approach to hybrid materials that combine excitons of different dimensionality could enable materials-level control of optical coupling through composition, nanoscale spacing, and spectral detuning.\par


Cross-dimensional dipole coupling is particularly attractive in hybrid excitonic materials that combine zero-dimensional (0D) and one-dimensional (1D) building blocks. Previous architectures used carbon nanotube (CNT)–chromophore and nanotube–quantum-dot hybrids and have mainly focused on localized donor–acceptor complexes, charge transfer, or Förster-type energy transfer effects. \cite{Ernst2012, Shafran2010} but these approaches often require chemical functionalization or are constrained by molecule size. In contrast, extended hybrid films offer a platform in which the optical response can be tuned through the composition, nanoscale proximity, and spectral detuning of the constituent excitons. Here, we assemble InP/ZnS quantum dots (QDs) and metallic CNTs into compact films by vacuum filtration.\cite{Wroblewska2021} The QD ensemble provides a comparatively fixed excitonic resonance, whereas the chirality-dependent CNT transitions span a broad spectral range. This combination enables multiple CNT excitons to probe their interaction with the same QD resonance within one material platform. Moreover, the distinct Raman fingerprints of the QDs and CNTs allow the optical response of each constituent to be addressed separately, providing a direct route to study cross-dimensional dipolar coupling in a multicomponent excitonic film.\cite{Simpson2018}\par

Optical spectroscopy of multicomponent excitonic materials is often complicated by overlapping transitions and ensemble inhomogeneity, which can confuse spectrally overlapping response of an individual constituent in absorption or scattering measurements. \cite{Manjavacas2011, Rodel2022, Davoodi2022} Energy-transfer measurements provide complementary information, but typically report population transfer from an initially excited donor to an energetically accessible acceptor state.\cite{Prasai2015, Ernst2012, Li2015, Rodarte2017}  Resonant Raman spectroscopy (RRs) can offer a different appoach where excitonic transitions are probed through crystal lattice-specific vibrational modes, enabling the response of individual components to be distinguished within a hybrid material.\cite{Araujo2007,Doorn2008, Gordeev2017, Yu1995, _sectionCardona1982} In CNTs the radial-breathing modes (RBM) provide chirality-selective fingerprints and allow the corresponding excitonic transitions to be mapped through Raman excitation profiles.\cite{Maultzsch2005a} Dipolar interactions between excitons can be described using coupled-oscillator models, which have been widely applied to coupled phonons\cite{Barker1964,scott1970, Hill2023} and exciton–plasmon systems.\cite{Thomas2018, Mueller2021} However, how such interactions renormalize the Raman-scattering process remains unexplored.\par

This work investigates how cross-dimensional dipole–dipole interactions modify resonant Raman scattering in quantum-dot–carbon-nanotube hybrid films. We develop a theoretical framework that combines a coupled-oscillator description of QD–CNT exciton interactions with perturbation theory for Raman scattering. In this approach, dipolar coupling renormalizes the exciton–photon matrix elements and produces detuning-dependent changes in the Raman excitation profiles. The model predicts a redshift of CNT resonances below the QD exciton energy, a blueshift of resonances above it, and near-resonant intensity redistribution and two-branch behavior. We test these predictions in films containing InP/ZnS QDs and metallic single-walled CNTs with chirality-dependent excitonic transitions spanning the QD resonance. The measured Raman excitation profiles follow the predicted detuning dependence and allow us to quantify an effective QD–CNT coupling energy.\par




\section{Results and discussion}

\subsection{Theory of dipole-coupled Raman scattering}

We model the optical response of a chirality-resolved CNT exciton coupled to the QD ensemble as two driven, damped harmonic oscillators. In this representation oscillator coordinates are proportional to the exciton transition dipole moments of the CNT ($\mu_{\mathrm{CNT}}$) and QD excitons ($\mu_{\mathrm{QD}}$). Both dipoles are driven by the exterior electric field $E_0cos(\omega t)$ and couple with dipole-dipole interaction term $M_c$. Their coupled dynamics is described by:\cite{Thomas2018, Mueller2021}
\begin{equation}
 \begin{alignedat}{2}
  \frac{d^2 \mu_{QD}(t)}{d t^2} + \omega^2_{QD} \mu_{QD}(t) +\gamma_{QD} \frac{d \mu_{QD}(t)}{d t} =A_{QD}\left[E_0 cos(\omega t) + M_c \mu_\mathrm{CNT} \right], \\  
   \frac{d^2 \mu_\mathrm{CNT}(t)}{d t^2} + \omega^2_\mathrm{CNT} \mu_\mathrm{CNT}(t) +\gamma_\mathrm{CNT} \frac{d \mu_\mathrm{CNT}(t)}{d t} =A_\mathrm{CNT}\left[E_0 cos(\omega t) + M_c \mu_{QD} \right],
 \end{alignedat}
  \label{EQ:Kin}
\end{equation}
here $X $ represents either $ CNT$ or $QD$. $A_X$ characterizes the oscillator strength, $\omega_X$ is the exciton resonance frequency, and $\gamma_X$ is the damping rate of the corresponding optical transition. The driving frequency is $\omega$ is defined by the laser energy $\hbar\omega = E_{las}$. This established model captures the linear optical response of two dipolar excitations and provides the starting point for incorporating dipole coupling into the Raman-scattering amplitude.\par
\begin{figure}
  \centering
  \includegraphics[width= 15cm]{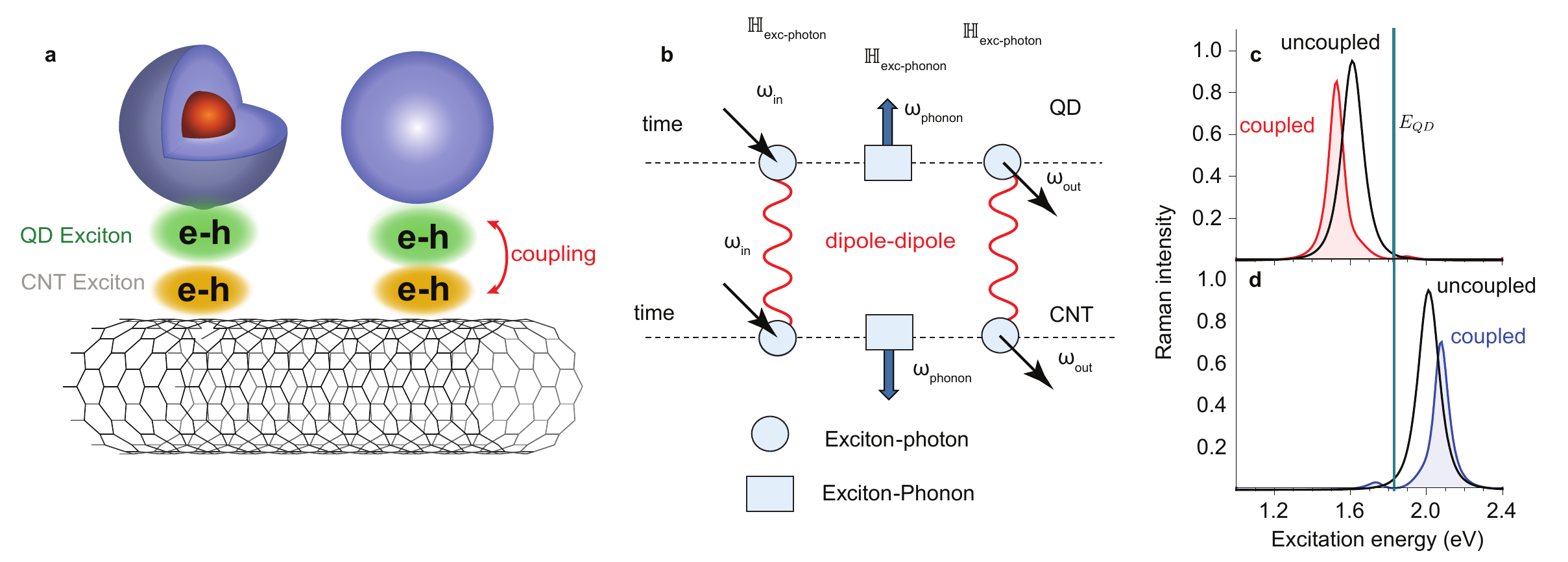}
  \caption{Dipole-dipole coupling in Raman scattering. (a) Schematic diagram demonstrating dipolar interaction of QD electron-hole pair with CNT electron-hole pair (b) Feynman diagram of Raman scattering processes with dipole-dipole coupling included. The tilted arrows represent photons ($\omega_{in},\omega_{out}$), where interaction with excitons occurs at round vertices. The vertical blue arrows and rectangular vertices indicate the interaction with the phonon. The red line interconnects the QD and CNT excitons with dipole-dipole coupling. (c) Resonant Raman profiles computed with Eq. \eqref{EQ:RCst} when $E_\mathrm{CNT}<E_{QD}$, set as 1.6 and 1.8 eV, respectively, black line is uncoupled ($M_c=0$) and red line is coupled system ($M_c=10$ eV). Panel (d) is the same as (c) for $E_\mathrm{CNT}>E_{QD}$, with $E_\mathrm{CNT} = 2$eV. 
  }
  \label{FIG:sch}
\end{figure}

Solution of Eq. \eqref{EQ:Kin} yields the driven CNT dipole response in the presence of the coupled QD excitation  in the steady-state regime

\begin{equation}
  \mu^0_\mathrm{CNT}(\omega)=\frac{E_0 A_\mathrm{CNT} \left[ 1 + \frac{A_{QD}}{B_{QD}(\omega)} \right] }{B_\mathrm{CNT}(\omega)\left[1 - \frac{A_{QD} A_\mathrm{CNT} M^2_c}{B_{QD}(\omega)B_\mathrm{CNT}(\omega)} \right] } =E_0\frac{A_\mathrm{CNT}^  {eff} (\omega)}{B_\mathrm{CNT}(\omega)}
  \label{EQ:Mom}
\end{equation}
where the $B_X = \hbar \omega_X^2 - \hbar\omega^2-i \hbar\omega \gamma_X $. Equation \eqref{EQ:Mom} defines an effective frequency-dependent CNT response, $\mu_{\mathrm{CNT}}^{0}(\omega)$, in which the QD resonance modifies both the amplitude and spectral shape of the driven CNT polarization. In the uncoupled limit, $M_c\rightarrow0$, this expression reduces to the conventional response of an isolated CNT exciton. We therefore define an effective CNT oscillator-strength factor, $A_{\mathrm{CNT}}^{\mathrm{eff}}(\omega)$, through Eq. \eqref{EQ:Mom}. This frequency-dependent optical response is subsequently incorporated into the exciton–photon matrix elements entering the Raman-scattering amplitude.\par

The frequency-dependent response $A_{\mathrm{CNT}}^{\mathrm{eff}}(\omega)$ is proportional to oscillator strength $f_{\mathrm{CNT}}^{\mathrm{eff}}$ in a Lorentz oscillator model.\cite{Thomas2018} It renormalizes simultaneously the optical excitation and emission vertices of the CNT exciton. We therefore express the effective exciton-photon matrix element as \cite{Yu1995}
\begin{equation}
 M_{\mathrm{ex-photon}}^{\mathrm{eff}}(\hbar\omega)
\propto
\frac{m_e}{e\sqrt{\hbar\omega_{\mathrm{CNT}}}}
\sqrt{f_{\mathrm{CNT}}^{\mathrm{eff}}(\hbar\omega)} \propto \sqrt{A_{\mathrm{CNT}}^{\mathrm{eff}}(\hbar\omega).}
\end{equation}
Unlike the bare oscillator strength, this effective matrix element is energy dependent because of the dipolar interaction with the QD resonance. It is this energy dependence that produces the characteristic modifications of the resonant Raman excitation profiles.\par

A typical one-phonon Raman process occurs in three steps, as represented by the top line of the Feynman diagram in Figure \ref{FIG:sch}b. First, the photon $E_L$ is absorbed by pushing the exciton from the ground into the excited state.  The interaction is described by an exciton-photon Hamiltonian, the round vertex in Figure \ref{FIG:sch}. In the second step, a phonon is excited within the Stokes Raman process, and the exciton is scattered into a new state. Finally, the exciton relaxes back to the ground state by emitting a photon of smaller energy $E_s = E_{in}-\hbar \omega_{ph}$. Without dipole-dipole interaction of the excitons, the Raman scattering cross-section can be represented by the standard third-order perturbation theory:\par
\begin{equation}
    I_{Ram}(\hbar\omega) \propto (\hbar\omega)^4{\left| \sum_{i}
      \frac{ M_{i,ex-photon}^{incident}  M_{i,ex-ph}M_{i,ex-photon}^{scattered}}{D^i_{in}(\hbar\omega)D^i_{out}(\hbar\omega)} 
    \right| }^2
    ,
    \label{EQ:RCst}
\end{equation}
where (i) labels the excitonic intermediate states. The sum runs over the exciton states with energies $\hbar \omega_{i}$. The $D_{in,out}$ represent the resonant enhancement by the incoming and outgoing Raman resonances $D^i_{in}(\omega)= (\hbar\omega-\hbar \omega_{i}-I\hbar \gamma_{i}/2)$ and $D^i_{out}(\omega) = (\hbar\omega-\hbar\omega_{ph}-\hbar \omega_{i}-I\hbar \gamma_{i}/2)$. The exciton-phonon $M_{i,ex-phonon}$ and exciton-photon matrix elements $M_{i,ex-photon}$ are independent of excitation energy $\omega$ without the dipole-dipole coupling. The experimentally measured profiles are normalized to a Raman scatterer to remove the trivial $E_L^4$ dependence.\par



We incorporate the dipole-dipole coupling effect into the Raman scattering formalism by replacing the bare exciton-photon matrix elements in Eq. \eqref{EQ:RCst} with the energy-dependent matrix element $M_{ex-photon}^{eff}(E_{las})$. Because the Raman process contains both photon absorption and photon emission, the coupled response enters at the incoming energy $E_L$ and at the outgoing energy $E_S=E_L-E_{\mathrm{ph}}$. For a CNT exciton coupled to the QD ensemble, the Raman excitation profile becomes
\begin{equation}
    I_{Ram}^{dipole-dipole}\propto \hbar \omega^4{\left| 
      \frac{ A_{eff}^{1/2}(\hbar \omega) A_{eff}^{1/2}(\hbar \omega-E_{ph})}{D_{in}(\hbar \omega)D_{out}(\hbar \omega)}   \right| }^2.
    \label{EQ:RRdd}
\end{equation}

Here $A_{\mathrm{CNT}}^{\mathrm{eff}}(E)$ is obtained from Eq. \eqref{EQ:Mom} and describes the coupling-renormalized optical response of the CNT exciton. In the limit $M_c\rightarrow0, A_{\mathrm{CNT}}^{\mathrm{eff}}$ becomes energy independent and Eq. \eqref{EQ:RCst} reduces to the conventional Raman expression in Eq. \eqref{EQ:RRdd}. The QD resonance therefore modifies the Raman excitation profile not by introducing an additional phonon-scattering pathway, but by changing the optical excitation and emission amplitudes of the CNT virtual intermediate state. \par


Our theory predicts a detuning-dependent displacement of the CNT Raman excitation profile, as illustrated in Figure \ref{FIG:sch}c,d. When the CNT exciton lies below the QD resonance, $E_{\mathrm{CNT}} < E_{\mathrm{QD}}, $ the coupled Raman profile shifts toward lower excitation energies relative to the uncoupled CNT response. Conversely, for
$ E_{\mathrm{CNT}} > E_{\mathrm{QD}}, $ the profile shifts toward higher energies. In both cases, the QD resonance can introduce an additional spectral feature near $ E_{\mathrm{QD}}, $ reflecting the QD-mediated renormalization of the CNT optical excitation and emission amplitudes. The magnitude of the shift, the spectral-weight redistribution, and the visibility of the QD-related feature depend on the coupling strength, oscillator strengths, damping rates, and excitonic detuning in Eq. \eqref{EQ:RRdd}. Thus, chirality-resolved Raman excitation profiles provide a direct means to track the sign and magnitude of the effective QD-CNT interaction.\par


When the CNT and QD excitons are nearly resonant, their interaction can produce two hybridized branches, denoted $E_+$ and $E_-$. These branches modify the Raman excitation profile through the incoming and outgoing resonance conditions in Eq. 5. Depending on the relative magnitude of the coupling-induced splitting and the damping rates, the hybrid response may appear as two resolved Raman resonances or as an asymmetric and broadened excitation profile. The energy separation between the hybrid branches provides can be expressed as
effective normal-mode splitting, $2\Omega$, obtained from the coupled-oscillator parameters as
\begin{equation}
  2\Omega = \frac{M_c\sqrt{A_{CNT}A_\mathrm{QD}}}{\hbar \omega_\mathrm{QD}}.
  \label{EQ:Rab}
\end{equation}
and provides an independent estimate of the effective QD-CNT coupling strength. In the following, we use both the detuning-dependent Raman-profile shifts and the near-resonant branch separation to quantify the coupling in the hybrid films.\par

\begin{figure}
    \centering
    \includegraphics[width = 6cm]{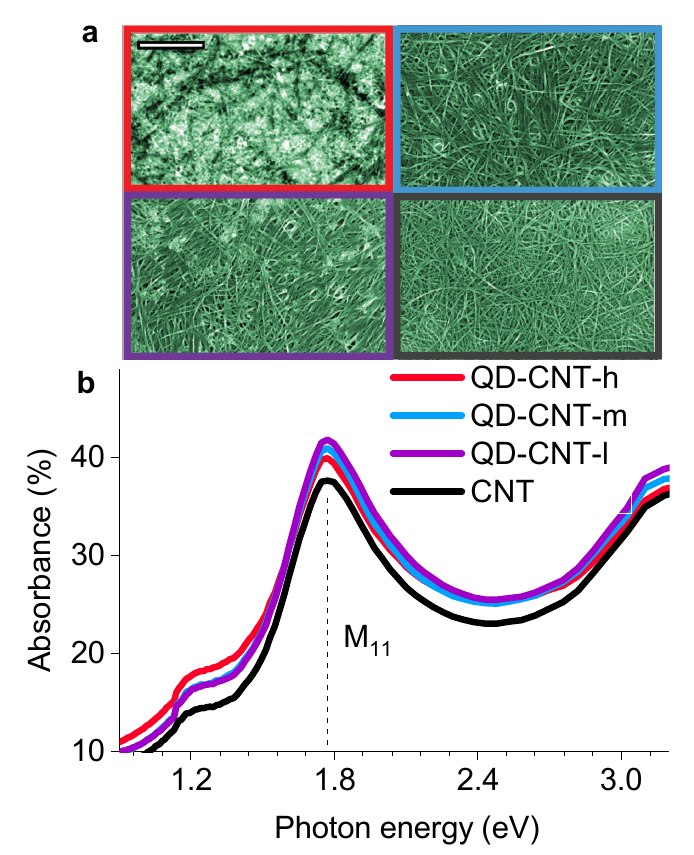}
    \caption{Pre-characterization of the CNT-QD films. (a) SEM images from hybrids at different QDs concentrations, increasing from pristine CNTs (black) to the QD-CNT-h (red). (b) Optical absorption spectra of films deposited on a glass substrate. }
    \label{FIG:PreCh2}
\end{figure}
\subsection{Pre-characterization}

We experimentally probe the dipole-dipole coupling effect using hybrid films with QDs and CNTs. The coupling is tuned by changing the relative concentration of QDs and CNTs during the vacuum filtration process. We incrementally increase the QD concentration from low to high and monitor the surface of the films by scanning electron microscopy (SEM). Figure \ref{FIG:PreCh2} a shows at the bottom the pristine films without QDs. The film contains randomly oriented bundles of nanotubes, similar to previous works.\cite{Wroblewska2021} By adding a small volume of QDs we start to observe white dotted areas on the surface of the CNT-QD-l sample, see Figure \ref{FIG:PreCh2}a (light blue). When further increasing the concentration of QDs up to medium and high, CNT-QD-m, h, more and more quantum dots appear in the SEM images, see Figure \ref{FIG:PreCh2}a (purple, red). It is important to keep the filtration speed low in order to eliminate the residues of the QD ligands, as for example shown in Supplementary Figures 3 and 4. The resulting films are $\approx$ \SI{50}{nm} thick as we estimated by atomic force microscopy.\par 



The linear optical spectra provide only limited evidence for QD-CNT coupling. Figure \ref{FIG:PreCh2}b compares the absorption spectra of pristine CNT films and QD-CNT hybrid films with increasing QD loading. In all samples, the optical response is dominated by the CNT excitonic absorption band centered near \SI{1.8}{\eV}. Incorporation of QDs produces only modest changes in the absolute absorbance and spectral line shape, despite the increasing QD density observed by SEM. In particular, no clearly resolved splitting or systematic shift of the broad CNT absorption feature is observed. This weak response reflects the overlap of multiple CNT chiralities and the ensemble nature of the absorption measurement. We therefore turn to resonant Raman spectroscopy, which resolves individual CNT excitonic transitions through their chirality-specific vibrational fingerprints and is consequently more sensitive to the QD-CNT dipolar interaction.\par

\begin{figure}
    \centering
    \includegraphics[width = 8cm]{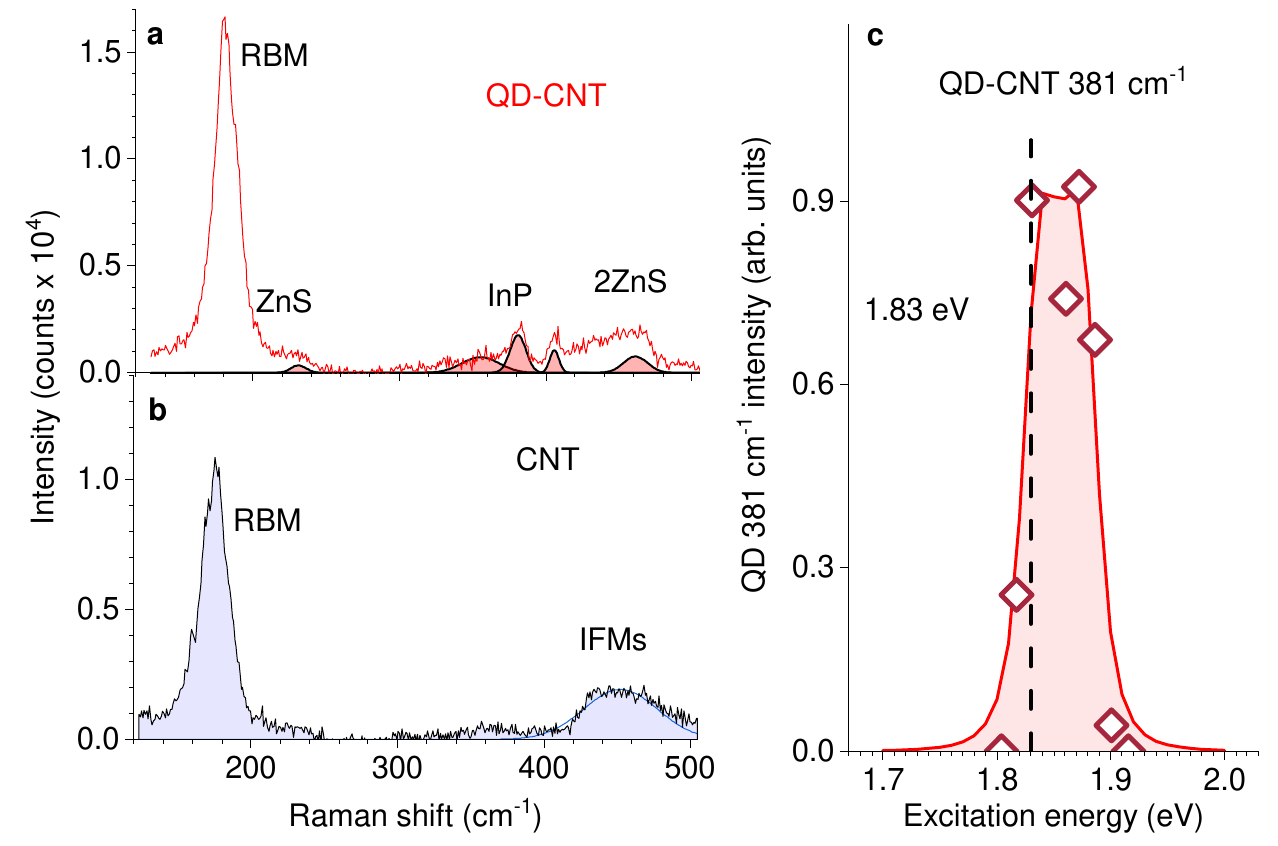}
    \caption{Resonant Raman analysis of the QD-CNT hybrid films  (a) Raman spectrum of QD-CNT hybrid film excited in resonance with quantum dots at 1.87 eV. The Raman modes of QDs are highlighted in red. (b) Raman spectrum of the pristine CNTs. The CNT Raman peaks are highlighted in blue. (c) Resonant Raman profile of the 381 cm$^{-1}$ Raman mode belonging to the QDs. Symbols are experimental data, line is fit with Eq. \eqref{EQ:RCst}. The vertical line indicates the energy of the optical transition ($E_\mathrm{QD}=1.83 eV , \gamma = 20$ meV). }
    \label{FIG:PreCh}
\end{figure}

Raman spectroscopy resolved distinct vibrational signatures of the InP/ZnS qunatum dots inside the hybrid films. Figure \ref{FIG:PreCh}a,b compares Raman spectra acquired with \SI{1.9}{eV} excitation from QD-CNT hybrid and pristine CNTs.  In the pristine film, we find two sets of modes: RBM between 150 and 200 cm$^{-1}$ and broad intermediate frequency modes (IFM) near 450 cm$^{-1}$, see Figure \ref{FIG:PreCh}b. The radial breathing modes represent an ensemble of peaks and we analyze them in the next paragraphs. In the Raman spectrum of the QDs-CNT hybrid additional Raman modes appear; see the highlighted peaks in Figure \ref{FIG:PreCh}a. The first (second)- order Raman modes of ZnS shells is found at $\approx$ 225 (450) cm$^{-1}$. These frequencies are comparable with previous studies, acquired at cryogenic temperatures.\cite{Brodu2018, Rafipoor2019} The Raman modes of InP cores are present at 355, 381, and 405 cm$^{-1}$, up-shifted compared to pure InP QDs.\cite{Seong2003} This is expected for core/shell QDs.\cite{Brodu2018}. Overall, the QD and CNT provide distinct, spectrally separable vibrational signatures to the hybrid films.\par

By varying laser excitation energy we can determine the energy of QD exciton $\hbar \omega_\mathrm{QD}$ inside the hybrid films. We quantify the energy of QDs exciton by resonant Raman spectroscopy. The Raman peak at \SI{381}{\per \cm} belonging to the InP shells is used as a spectroscopic marker of the QD response, see Figure \ref{FIG:PreCh}c. We record the Raman spectrum at several excitation energies between 1.75 and \SI{1.95}{eV} and fit the InP Raman peak with a Lorentzian function. The Raman intensity is divided by the Raman intensity of a benzonitrile solution to reference out the sensitivity of the spectrometer; see Methods. The calibrated intensity is plotted as symbols in Figure \ref{FIG:PreCh}c. Far from the resonance at 1.8 and \SI{1.9}{eV} we find faint Raman intensity, while the maximum is reached in between the incoming and outgoing resonances at $\hbar \omega_\mathrm{QD}$ and $\hbar \omega_\mathrm{QD}+E_{vib}$, respectively. We fit the experimental data by Eq. \eqref{EQ:RRdd} with  $\hbar \omega_\mathrm{QD}$ as free parameter, see lines in Fig. \ref{FIG:PreCh}c, and obtain the $E_\mathrm{QD}$ = \SI{1.83}{eV}. Note that this energy is red-shifted from individual QDs, altered by dielectric screening \cite{Elamathi2020} and clustering effects\cite{Vossmeyer1994, Micic1998}. In the following, $E_{\mathrm{QD}}$ is used as the fixed QD resonance energy in the analysis of the CNT Raman excitation profiles.\par
 
 \begin{figure}
     \centering
     \includegraphics[width=8cm]{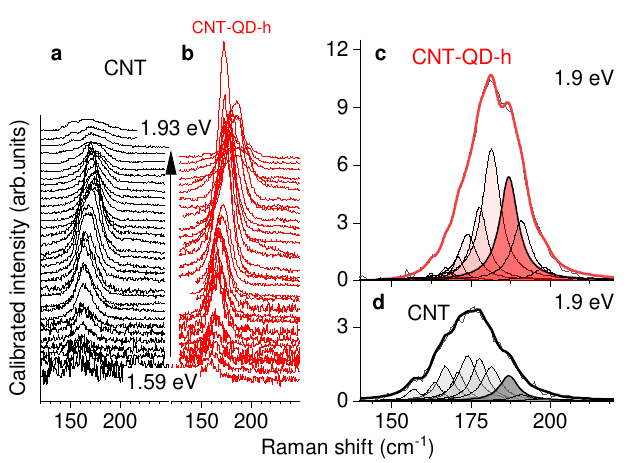}
     \caption{Analysis of the radial breathing Raman modes of CNTs. (a,b) Waterfall plot with calibrated intensity in (a) pristine and (b) QD-hybrid samples with excitation energies spanning from 1.56 (bottom) to \SI{2}{eV} (top). (c,d) Decomposition of the spectra excited at 1.9 eV intro separate chiralities, with filled peaks representing different (n,m) species. The highlighted peak represents the \SI{187}{\per\cm} RBM.}
     \label{FIG:CNT_RR}
 \end{figure}
\subsection{Optical properties of the CNTs}

We next record resonant Raman spectra of the CNT species in the pristine and hybrid films and identify individual chiralities, since mixed-chirality samples contain signals from several CNT chiralities (Fig. \ref{FIG:CNT_RR}a). Figures \ref{FIG:CNT_RR}b and c compare the Raman spectra of QD-CNTs-h and CNTs, respectively, both excited at \SI{1.9}{eV}. Although the overall RBM envelope differs between the two samples, the dominant effect is not a substantial shift of the RBM frequencies themselves, but a redistribution of intensity among individual RBM components. This behavior reflects altered resonance conditions for different nanotube chiralities after incorporation of QDs. \cite{Maultzsch2005a}. We break down the Raman spectrum into individual RBM peaks according to the Kataura plot \cite{Maultzsch2005a,Gordeev2017}, with shown in Figure \ref{FIG:CNT_RR}b,c. The relative weights of these components differ between the pristine and hybrid films, indicating that the QD-induced interaction selectively modifies the resonant Raman response of CNT excitons with different transition energies. \par

\begin{figure}
    \centering
    \includegraphics[width =16.5cm]{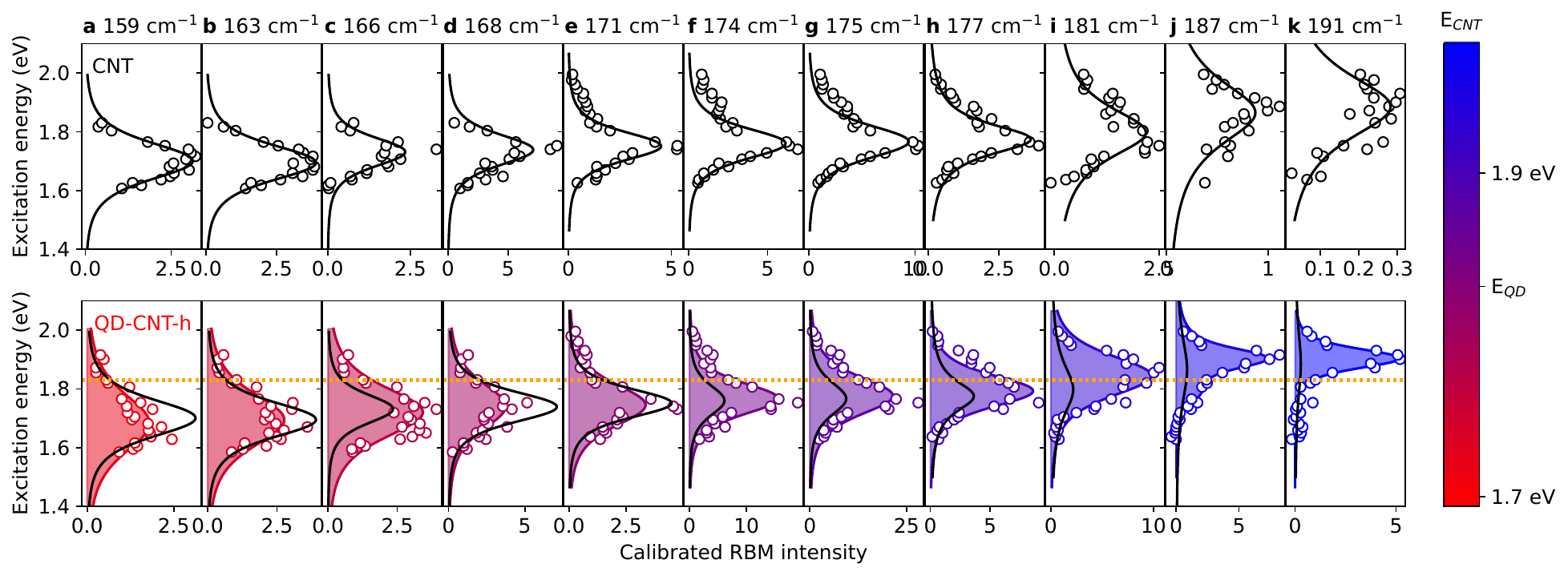}
    \caption{Resonant Raman profiles of the pristine and hybrid CNTs, with RBM frequencies indicated on the top. Top row with pristine CNTs, the symbols are experimental data and the lines are fits with Eq.\eqref{EQ:RCst}. In the bottom row RR profiles of pristine CNTs (black) are compared with QD-CNT-h (red-blue), the lines are fits with Eq. \eqref{EQ:RRdd}. The orange horizontal line represents the QD exciton energy $E_\mathrm{QD}$. }
    \label{FIG:RR_profiles}
\end{figure}

We apply the same spectral decomposition to the full excitation-energy series and construct chirality-resolved Raman excitation profiles for the individual RBM components, for pristine top panels in Figure \ref{FIG:RR_profiles} and QD-CNT-h samples (bottom panels in Figure \ref{FIG:RR_profiles}). For CNT excitons below the QD resonance, the hybrid films exhibit a systematic displacement of the Raman profile toward lower excitation energies. First, we examine the resonant Raman profile of the \SI{159}{\per\cm} RBM mode in the pristine sample, shown in Figure \ref{FIG:RR_profiles}a in the top panel. Using a fit with Eq. \ref{EQ:RCst} we obtain the exciton energy of \SI{1.69}{eV}. The RR profile of the hybrid system is shown in the bottom panel of Figure \ref{FIG:RR_profiles}a. We observe a red-shift of the resonance and a broadening, with the $E^{-}$ energy at \SI{1.65}{eV}. The resulting shift of approximately \SI{38}{\meV} is consistent with the detuning-dependent response predicted for $ E_{\mathrm{CNT}} < E_{\mathrm{QD}}$. This comparison provides the first experimental indication that the QD resonance modifies the Raman response of off-resonant CNT excitons in the hybrid film.\par


The coupling-induced modification becomes most pronounced when the CNT and QD excitons are nearly resonant, as observed for the Raman profiles in Figures \ref{FIG:RR_profiles}h-k. We focus on the RBM mode at \wn{187} in Figure \ref{FIG:RR_profiles}j, where the resonance is broad and displays low intensity in the pristine film. We find the resonance energy of the pristine film at \SI{1.86}{eV} close to $E_\mathrm{QD}$ = \SI{1.83}{eV} and therefore pronounced interaction is expected. The RR profile of the hybrid film is shown in the bottom panel of Figure \ref{FIG:RR_profiles}j and is up to 7.5 times higher integrated Raman spectral weight compared to pristine films. The fits by Eq. \eqref{EQ:RRdd} nicely reproduce the asymmetric line shape of the resonance. The model parameters are $A_\mathrm{QD}=$ \SI{0.15}{eV\squared}, $A_{CNT}=$ \SI{0.096}{eV\squared}, and $M=0.85$. We plug these parameters in Eq. \eqref{EQ:Rab} and obtain a Rabi splitting $2\Omega$ = 110meV. The CNTs exciton splits into two parts $E^+$ and $E^-$ at \SI{1.88}{eV} and for $E^-$ is \SI{1.76}{eV}, as determined from a fit with Eq. \eqref{EQ:RRdd}. The difference of the splitting energies $E^+- E^-$ is \SI{120}{meV} which almost coincides with the value obtained from the full perturbation approach in Eq. \eqref{EQ:RRdd}. The close agreement between this separation and the value obtained from the coupled-oscillator parameters supports the assignment of the near-resonant Raman profile to QD--CNT dipolar coupling.\par

 
 \begin{figure}
     \centering
     \includegraphics[width=8cm]{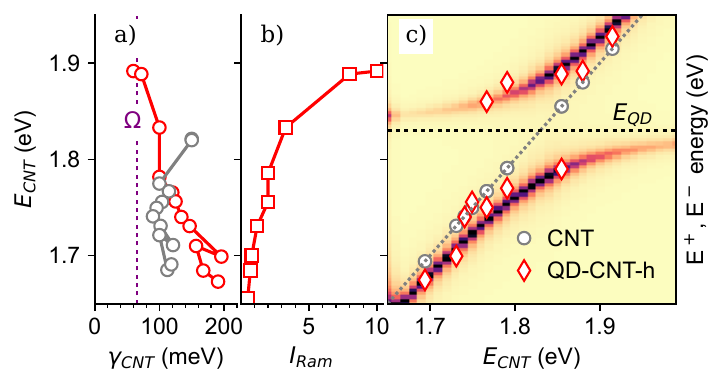}
     \caption{Effects of dd coupling between QD and CNT excitons on CNT exciton energy and broadening. (a) Spectral broadening ($\gamma_{CNT}$) extracted from the resonant Raman profiles. (b) Raman enhancement calculated as a ratio between spectral weights of resonant Raman profiles $\frac{P_{QD-CNT}}{P_CNT}$.  (c) Rabi splitting in the dipole-dipole coupling model. The circles represent $E_{CNT}$ in the pristine system and diamonds the $E^{\pm}$ in CNT-QD-h hybrid films. Dashed lines represent the energies of the uncoupled QD and CNTs.}
     \label{FIG:F4}
 \end{figure}

The chirality-resolved Raman excitation profiles reveal systematic changes in both Raman efficiency and effective spectral width as the CNT exciton energy approaches the QD resonance. The RR profiles of the 11 pristine CNT chiralities are summarized in Figure \ref{FIG:RR_profiles} (top row). The lines represent fits by Eq.\eqref{EQ:RCst}, the resonance maximum shift together with exciton resonance energies $E_{CNT}$. $E_{CNT}$ increases with the RBM frequency as expected from the Kataura plot. The higher $E_{CNT}$ the higher is the profile broadening, consistent with the previous observations in metallic CNTs. The RR profiles of the QD-CNT-h hybrids are compared with fits of the pristine profiles in the bottom row in Figure \ref{FIG:RR_profiles}. For the lowest energy CNT with RBM at \SI{159}{\per \cm} we find a broader profile with an overall smaller amplitude, however in the highest energy \SI{191}{\per \cm} RBM we find up to 10 times stronger RR profile, compared to the pristine CNTs. It is nicely visible how the intensity increases and broadening decreases in the hybrid system from left to right in Figure \ref{FIG:RR_profiles}. The variation of the broadening factors and Raman efficiencies is summarized in Figure \ref{FIG:F4}a,b.

The QD-induced modification of the CNT Raman excitation profiles depends systematically on the excitonic detuning. WFor CNT excitons below the QD resonance, $E_{\mathrm{CNT}} < E_{\mathrm{QD}}$, we find an overall red-shift of the resonance (\ref{FIG:RR_profiles}, bottom). For CNT excitons above the QD resonance, $E_{\mathrm{CNT}} > E_{\mathrm{QD}}$, the shift reverses toward higher energies. As $E_{\mathrm{CNT}}$ approaches $E_{\mathrm{QD}}$, the Raman response evolves into two hybrid branches, $E_{+}$ and $E_{-}$. We fit the resonant Raman profiles from Figure \ref{FIG:RR_profiles} and determine $E_{+,-}$ energies. The branch energies extracted from fits to the profiles in Figure \ref{FIG:RR_profiles} are plotted in Figure \ref{FIG:F4}c as a function of the corresponding uncoupled CNT exciton energy. They follow an avoided-crossing-like dependence around $E_{\mathrm{QD}}$, in agreement with the coupled-oscillator calculation. The background map represents the calculated CNT dipole response, $|\mu_{\mathrm{CNT}}|$, for an effective splitting of $2\Omega = 110\ \mathrm{meV}$.  The fitted linewidths of the pristine CNT excitons are on average approximately $120\ \mathrm{meV}$ and increase for higher-energy CNT transitions, Figure \ref{FIG:F4}a. Because these damping rates are comparable to or exceed the extracted coupling energy, the hybrid films are best described as operating in a dissipative coupling regime rather than a fully resolved coherent strong-coupling regime. Nevertheless, the chirality-selective Raman excitation profiles retain clear signatures of QD-CNT dipolar coupling, including the detuning-dependent spectral shifts, intensity redistribution, and near-resonant two-branch response.


\section{Discussion}

Our results show that dipole-dipole coupling between QD and CNT excitons can be resolved through constituent-specific Raman excitation profiles in a mixed-dimensional hybrid film. In contrast to linear absorption, which averages over the broad distribution of nanotube chiralities, resonant Raman spectroscopy isolates individual CNT transitions through their RBM fingerprints. This selectivity allows the detuning-dependent evolution of the coupled response to be followed across multiple CNT excitons interacting with the same QD resonance. The present framework therefore extends coupled-oscillator descriptions beyond linear optical spectra by incorporating the coupling-induced renormalization of the optical vertices in the Raman-scattering amplitude.\par

The QD-CNT film architecture provides several routes to engineer the interaction further. The detuning can be controlled through the QD size and composition or through the selection of CNT chiralities, whereas the coupling strength can be tuned by the QD density, interparticle separation, and dielectric environment. Monochiral CNT films and QDs with narrower size distributions should reduce inhomogeneous broadening and enable a more direct comparison between the coupling energy and the excitonic damping rates. Coupling may also be enhanced by integrating the hybrid films into photonic or plasmonic resonators. For coherently coupled and similarly oriented dipoles, the collective coupling is expected to scale as $\sqrt{N}$, where $N$ is the number of participating dipoles. A systematic QD-density series would provide a direct test of this scaling in the present hybrid-film platform. More broadly, the Raman formalism developed here can be applied to other mixed-dimensional excitonic heterostructures, including  2D-1D and 2D-2D hybrids.\cite{Geim2013,Guo2022,Gordeev2021}


\section{Conclusions}

We introduce quantum-dot--carbon-nanotube hybrid films as a cross-dimensional excitonic platform for studying dipole-coupled optical excitations. By combining a coupled-oscillator model with perturbation theory for resonant Raman scattering, we show that QD--CNT dipolar interactions renormalize the optical vertices of the Raman process and thereby modify the excitation profiles of individual CNT excitons. Experimentally, the Raman profiles exhibit detuning-dependent redshifts and blueshifts relative to pristine CNT films, together with intensity redistribution, reduced effective linewidths, and near-resonant two-branch behavior. The coupled response yields an effective splitting of approximately $110\ \mathrm{meV}$. Because individual CNT transitions can be isolated through their RBM fingerprints, resonant Raman spectroscopy reveals coupling effects that remain weakly apparent in ensemble absorption spectra. These results establish component-selective Raman spectroscopy as a tool for probing and engineering dipolar interactions in mixed-dimensional excitonic nanomaterials.\par

\section{Methods}

Hybrid QD–CNT films were prepared by vacuum filtration. Metallic single-walled carbon nanotubes (NanoIntegris) were dispersed in an aqueous 1 wt \% sodium dodecyl sulfate (SDS) solution, sonicated for 5 h, and centrifuged at 8000 rpm for 10 min to remove large aggregates. InP/ZnS quantum dots (MkNano; stock concentration, 1 mg mL$^{-1}$) were diluted to 0.01 mg mL$^{-1}$. For each sample, 3.5 mL of the CNT dispersion was mixed with 0.5, 1.0, or 2.0 mL of the QD dispersion to produce the low-, medium-, and high-QD-loading films, denoted QD-CNT-l, QD-CNT-m, and QD-CNT-h, respectively. The mixtures were heated to 50 °C for 30 min and subsequently sonicated for 30 min.\par

The dispersions were filtered through cellulose membranes (Merck Millipore; diameter, 25 mm; pore size, 25 nm). The resulting films were transferred onto Si/SiO($_2$) substrates using an acetone-vapor treatment for 30 min, followed by immersion in warm acetone for 15 min and isopropanol for 3 min. The complete vacuum-filtration and transfer protocol is described in our previous work.\cite{Wroblewska2021} SEM imaging confirmed a higher density of QD-related contrast with increasing nominal QD loading, while atomic-force microscopy yielded film thicknesses of approximately 50 nm. Film morphology was characterized using a Raith e-Line Plus scanning electron microscope, and thicknesses were measured using a Bruker Icon atomic-force microscope.\par

Optical absorption spectra were acquired using a PerkinElmer LAMBDA 850 UV–vis–NIR spectrometer equipped with a 150 mm integrating sphere. Resonant Raman measurements were performed using the setup described previously.\cite{Wroblewska2021, Gordeev2017} Excitation was provided by a tunable Ti laser (Coherent MBR 110; 700–1000 nm), and the scattered light was analyzed with a triple-grating Horiba T64000 spectrometer equipped with a 900 lines per mm grating and a Peltier-cooled Si CCD detector. For each excitation energy, the Raman spectra were corrected for the spectral response of the detection system using a benzonitrile reference measurement. The resulting calibrated intensities were used to produce Raman excitation profiles with the QD and CNT vibrational modes.\par

\begin{acknowledgement}
A.W. acknowledges support by the Warsaw University of Technology within the Excellence Initiative: Research University (IDUB) program. S.R. acknowledges funding from the European Research Council (ERC) under Grant 772108. G.G. acknowledges funding from the Fond National de la Recherche under Project AWORD. N.S.M. acknowledges funding from the Deutsche Forschungsgemeinschaft (DFG, German Research Foundation) - Projektnummer 551280726.

\end{acknowledgement}

\begin{suppinfo}

Additional figures are present in the Supporting information.

\end{suppinfo}

\bibliography{main}

\end{document}


\renewcommand{\figurename}{Supplementary Figure}


\begin{abstract}

\end{abstract}

\tableofcontents

\section{Atomic force microscopy}
Atomic force microscope characterization was performed to check the thickness of the carbon nanotube layer. As a result, it was found that the tested material was about 50 nm thick. 

\begin{figure}
    \centering
    \includegraphics[width=9cm]{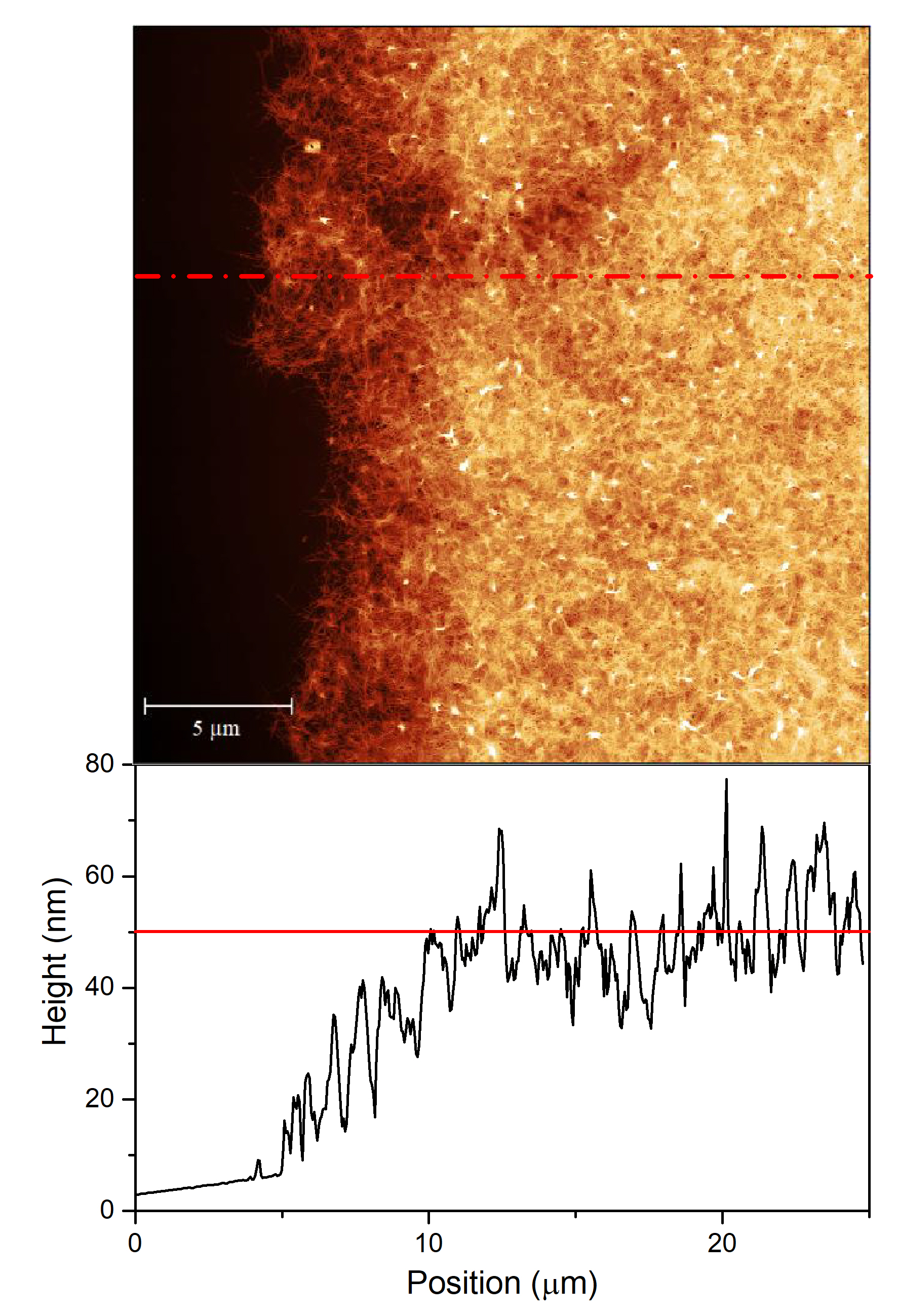}
    \caption{AFM image with cross section profile of the sample }
    \label{AFM_CNT}
\end{figure}

\newpage
\section{Scanning electron microscopy}
\begin{figure}[h]
    \centering
    \includegraphics[width= 13cm]{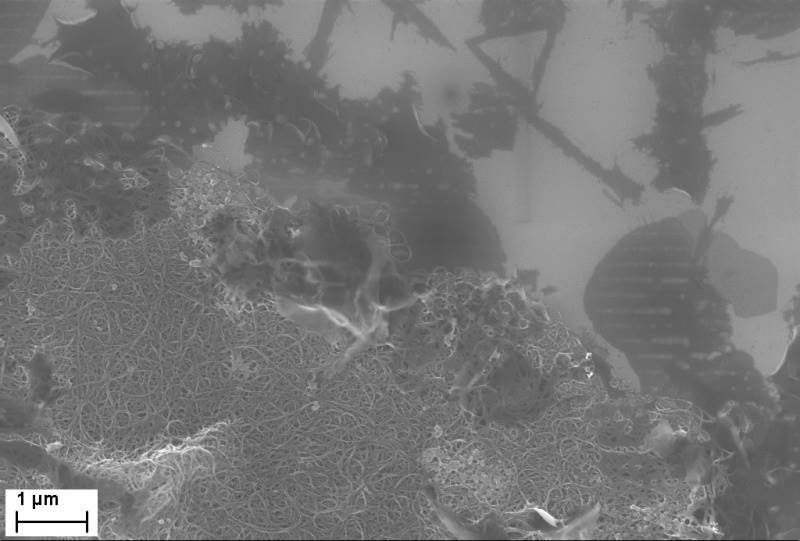}
    \caption{Thin film  of carbon nanotubes with InP/ZnS nanocrystals with surfactant residual}
    \label{SEM_impurities1}
\end{figure}

\begin{figure}
    \centering
    \includegraphics[width=13cm]{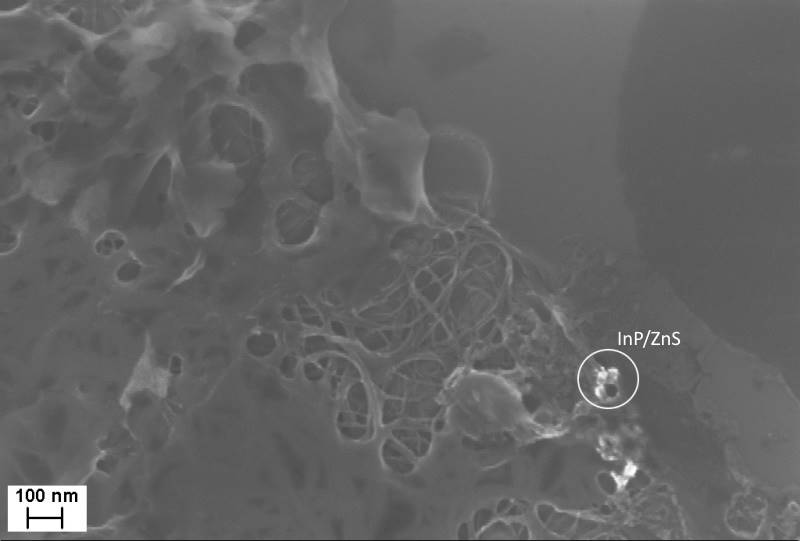}
    \caption{Thin film  of carbon nanotubes with InP/ZnS nanocrystals with surfactant residual}
    \label{SEM_impurities2}
\end{figure}

\begin{figure}
    \centering
    \includegraphics[width=13cm]{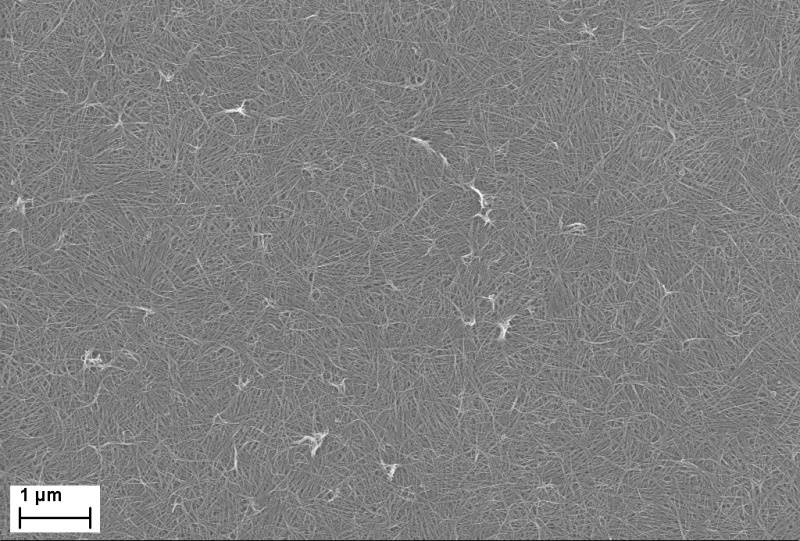}
    \caption{Thin film  of pure carbon nanotubes}
    \label{SEM_M-CNT}
\end{figure}

\begin{figure}
    \centering
    \includegraphics[width=13cm]{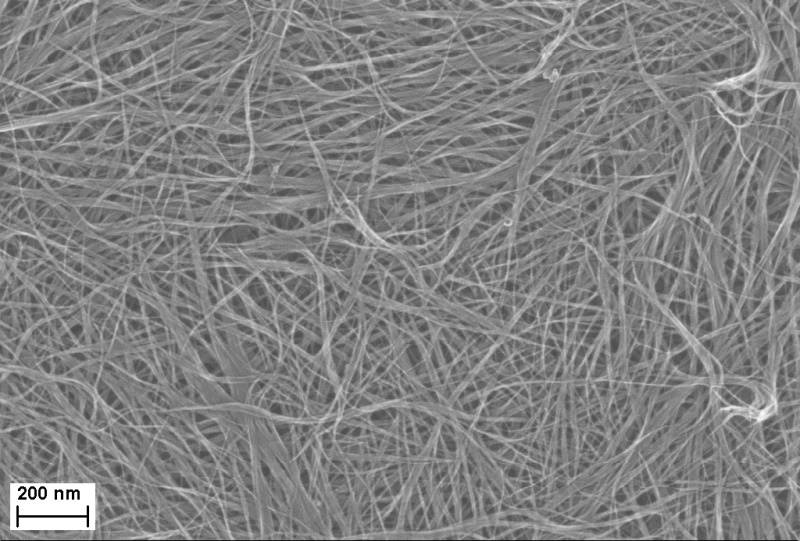}
    \caption{Thin film  of pure carbon nanotubes}
    \label{SEM_M-CNT2}
\end{figure}

The morphology of the layer surface plays an important role in the study of phenomena related to carbon nanotube nanotube-quantum dot interactions. The most important issue is to obtain clean layers, free of surfactants, catalyst residues, or residual reagents after the layer transfer process. To verify the surface of the sample, SEM images were taken. Figures \ref{SEM_impurities1} and \ref{SEM_impurities2} show thin films with carbon nanotubes and some contaminations. Fig. \ref{SEM_impurities2} shows labeled as InP/ZnS nanocrystal. In contrary, figures \ref{SEM_M-CNT} and \ref{SEM_M-CNT2} show a pure layer. In fig. \ref{SEM_InP} we present thin film with the highest concentration of InP/ZnS nanocrystals. 

\begin{figure}
    \centering
    \includegraphics[width=13cm]{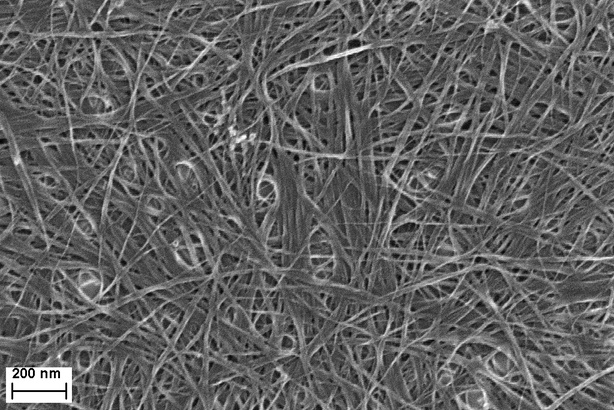}
    \caption{Thin film  of carbon nanotubes with lowest concentration of InP/ZnS nanocrystals}
    \label{SEM_InP}
\end{figure}

\begin{figure}
    \centering
    \includegraphics[width=13cm]{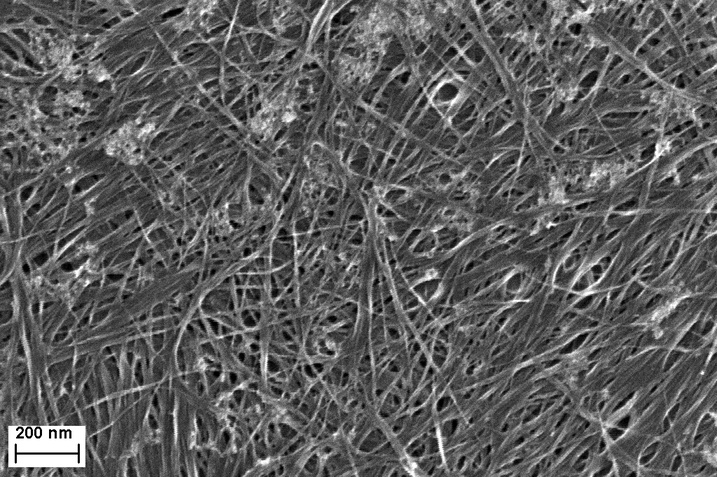}
    \caption{Thin film  of carbon nanotubes with medium concentration of InP/ZnS nanocrystals}
    \label{SEM_InP}
\end{figure}

\begin{figure}
    \centering
    \includegraphics[width=13cm]{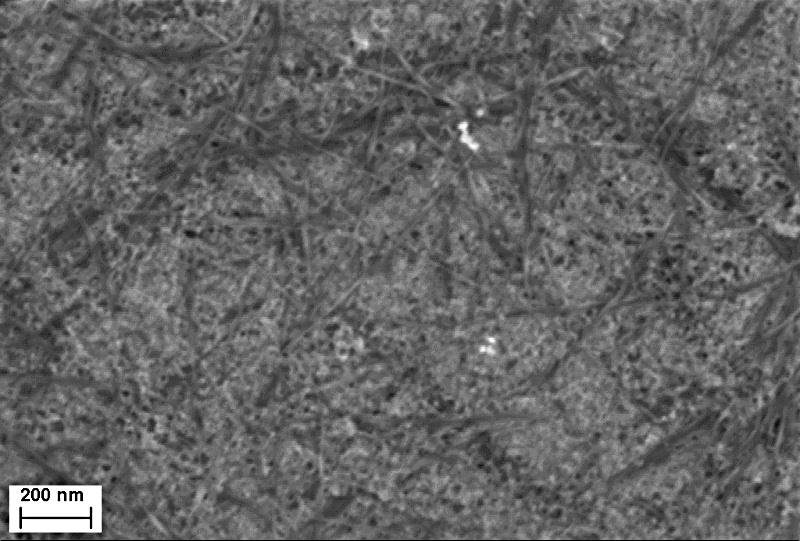}
    \caption{Thin film  of carbon nanotubes with highest concentration of InP/ZnS nanocrystals}
    \label{SEM_InP}
\end{figure}

   \newpage

 
\section{Raman doping analysis}

\begin{figure}
    \centering
    \includegraphics[width=9cm]{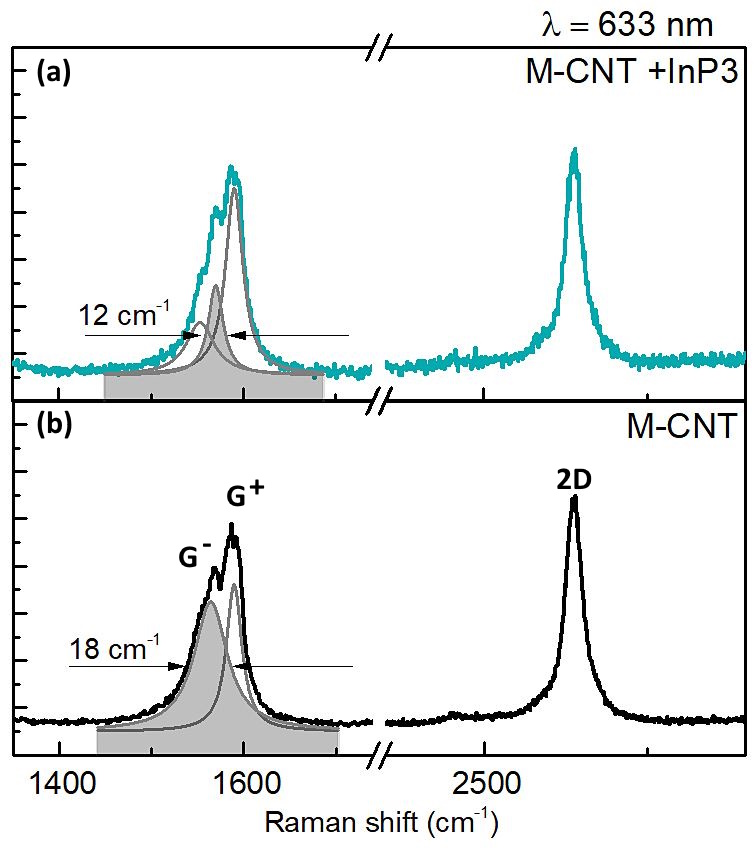}
    \caption{Raman spectrum for pristine metallic (black) and after InP/ZnS nanocrystals addition (green) using 633 nm laser line. The fitting peaks are marked and grey filled is analyzed LO phonon}
    \label{Raman_single}
\end{figure}

Figure \ref{Raman_single} shows example Raman spectra of pure CNT thin film (Fig \ref{Raman_single}a) and CNT thin films with embedded InP/ZnS nanocrystals (Fig. \ref{Raman_single}b). In the G band region, we can identify two main bands assigned as $G^{-}$ at smaller phonon frequency ($\approx$ 1584 cm$^{-1}$) and $G^{+}$ at larger phonon frequency ($\approx$ 1594 cm$^{-1}$). Bands correspond to two in-plane optical-phonon modes: longitudinal optical (LO) and transverse optical (TO) mode ($G^{-}$ and $G^{+}$, respectively). In metallic carbon nanotubes LO phonon is highly sensitive to doping effects due electron- phonon coupling (marked as grey area under the fitted curve).  In fig \ref{Raman_single}a and \ref{Raman_single}b we can observe a slight narrowing ($\sim$ 6 cm$^{-1}$) and shift into higher phonon energy ($\sim$ 2 cm$^{-1}$) of $G^{-}$ band. Based on these type of changes we can report changes in the doping level and might be a suggest of p type doping.

\begin{figure}
    \centering
    \includegraphics[width=14cm]{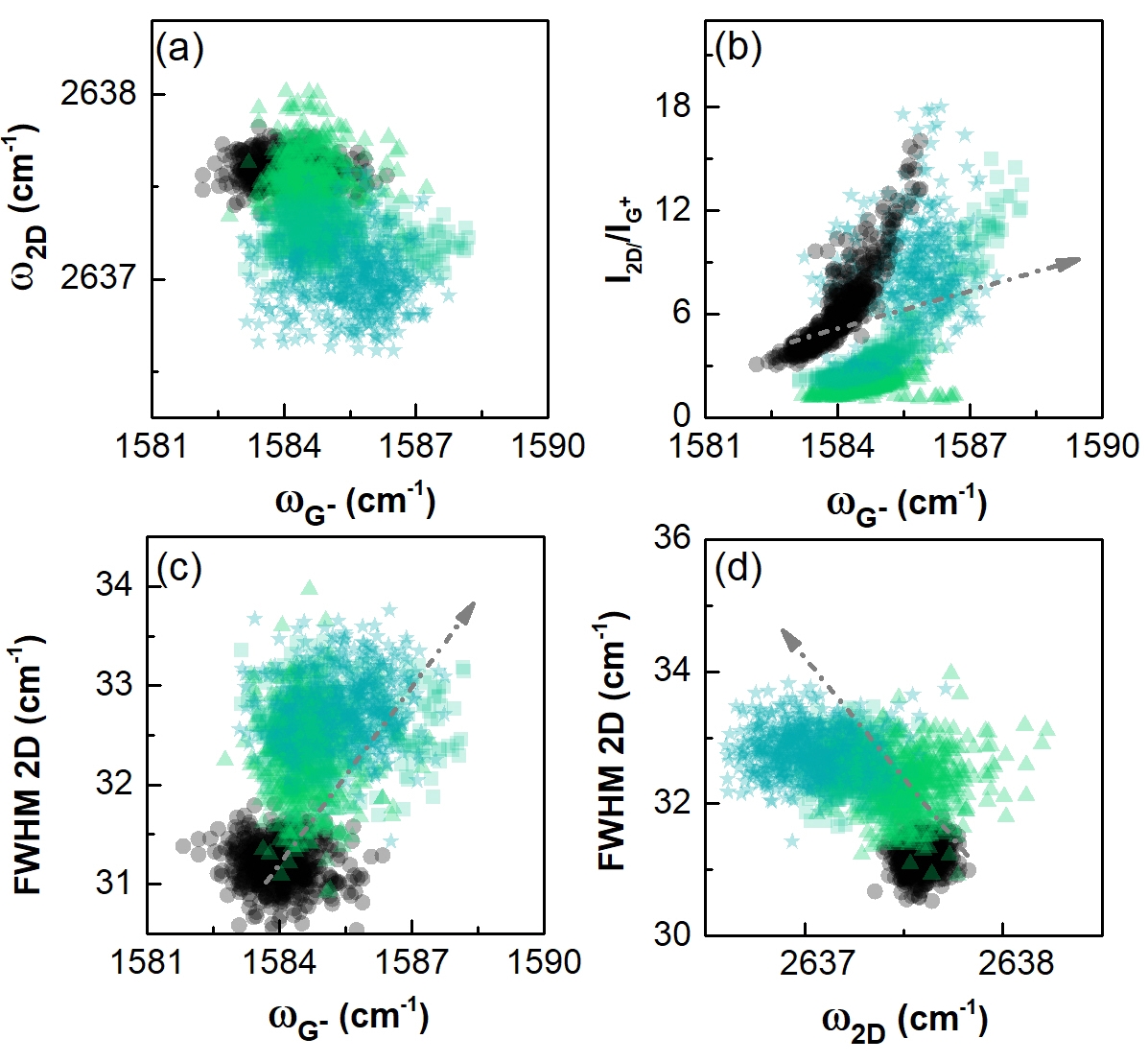}
    \caption{Correlations of the Raman peak parameters for metallic CNT thin films: pure (black) and with three concentrations of InP/ZnS QDs.}
    \label{Raman_correlations}
\end{figure}

Based on single Raman spectra, it is difficult to obtain satisfying information about global changes in the doping level or its type in layers containing nanoparticles in the whole volume. To study large areas of the samples, Raman maps and statistical analysis were used. In mapping mode, a 100 µm x 100 µm scan containing 600 individual spectra was performed. We used a low laser power ($\approx$0.35 mW) to avoid heating effects on the samples during the measurements. The measurements were performed at room temperature. Subsequently, the Gnuplot program was used to fit a Lorentz function to each band and obtain peak parameters: position, intensity and full width at half maximum. Results obtained from fitting procedure are presented in the form of correlation plots of each parameter. Fig. \ref{Raman_correlations} a) show the correlation of 2D band position plotted as a function of G- position. One can observe a shift of the 2D band toward lower energies as the G- band position increases. Based on the nature of these bands, we can conclude that p-type doping has been observed, which involves the transfer of electrons from the $\pi$ state of carbon nanotubes to nanocrystals generating holes in the electron structure of carbon nanotubes \cite{fouquet2009, ruch2009}. The study of the effect of InP/ZnS nanocrystals on the level of doping in metallic carbon nanotubes can also be determined from the I2D/IG- intensity ratio plotted as a function of the position of the G- band (Fig.  \ref{Raman_correlations}b)). An increase in the I2D/IG- intensity ratio was observed with increasing nanocrystal concentration. The observed effects are a result of the sensitivity of the 2D band to changes in the electron structure of the nanotubes. Furthermore, it can be seen that there is an effect of electron-phonon pair creation. The generation of this type of coupling has an impact on the intensity of the Raman bands due to the phenomenon of renormalisation of the phonon energy. The intensity of the 2D band increases with increasing electron density, which affects the value of the I2D/IG- ratio \cite{casiraghi2009doping, casiraghi2007raman}. Charge transfer or changes in the density of electron states of carbon nanotubes have a significant effect on the 2D FWHM (Fig \ref{Raman_correlations} c)). It was observed as the concentration of theInP/ZnS nanocrystals, FWHM 2D band increases and the G-band shifts towards higher energies. Using Raman maps with statistical analysis exposed the phenomenon of electron-phonon coupling, which is difficult to observe in single spectra. Description of the creation of electron-phonon coupling also results in renormalisation of the phonon energy, which affects the finite lifetime of phonons and electrons. Another explanation for the observed changes is the slight deformation of the structure, associated with the presence of nanocrystals throughout the layer \cite{das2009renormalization, tsang2007doping}. Based on the correlation of FWHM 2D band plotted dependent on position of this band (Fig. \ref{Raman_correlations}d)), one can determine whether the structure is affected only by doping, or if structure stress effects are also observed. It can be noted that the 2D band broaden with increasing concentration of InP/ZnS nanocrystals. This suggests presence of strain arising in the hybrid layer due the influence of nanocrystals \cite{cronin2004measuring, kumar2008optical}.

\bibliography{main}